# Conceptual Closure: How Memories are Woven into an Interconnected Worldview

LIANE M. GABORA

**ABSTRACT**

This paper describes a tentative model for how discrete memories transform into an interconnected conceptual network, or worldview, wherein relationships between memories are forged by way of abstractions. The model draws on Kauffman’s theory of how an information-evolving system could emerge through the formation and closure of an autocatalytic network. Here, the information units are not catalytic molecules, but memories and abstractions, and the process that connects them is not catalysis but reminding events (i.e. one memory evokes another). The result is a worldview that both structures, and is structured by, self-triggered streams of thought.

## 1. INTRODUCTION

Being a physically closed system, the living organism is richly interconnected (e.g., by way of sensorimotor and endocrine systems), which endows it with *behavioral contextuality*. That is, any perturbation will percolate through the interconnected system and elicit a response tailored to the specifics of the perturbation. Conceptual closure is a second level of closure, in conceptual space rather than physical space. That is, the mind is conceptually closed in the sense that every concept, belief, et cetera, impacts, and is impacted by, a sphere of related concepts and beliefs, and the network is closed in the sense that there exists a 'conceptual pathway' through streams of associative recall from any one concept to any other. This second form of closure further enhances the potential for behavioral contextuality by enabling the individual to engage in relational streams of associative thought that refine potential behaviors in light of goals or imagined outcomes. This paper addresses the question of how conceptual closure might be achieved. That is, how does the mind weave memories of specific experiences into a relationally-structured worldview? It is the relational or associative structure of the human mind that enables us to strategically generate and refine ideas, and thereby provides us with the capacity for culture. So the question is important.

The explanation proposed here was inspired by a theory of the origin of life. The origin of life and the origin of human cognition may appear at first glance to be very different problems.

However, deep down they amount to the same thing: the bootstrapping of a system by which variants of an information pattern are generated, and the selective proliferation of some variants over others [2, 7]. Culture, like biology, can be viewed as a form of evolution, albeit one that manifests differently from biological evolution. In keeping with this evolutionary framework, the term ‘meme’ will be used to refer to a unit of cultural information as it is represented in the brain. For our purposes, a meme can be anything from an idea for a recipe, to a memory of one’s uncle, to a concept of size, to an attitude of racial prejudice. The rationale for lumping these together is that they are all ‘food for thought’—units of information drawn upon to invent new memes or to clarify relationships amongst existing ones—that can be implemented as actions, vocalizations, or artifacts.

It is a good idea to start off by clarifying the difference between conceptual closure and psychic closure [22]. As I see it, the concepts are similar, but whereas psychic closure focuses more on individuals as intentional systems, conceptual closure focuses more on mental representations and their interrelations.

## 2. ORIGIN OF THE COGNITIVE UNDERPINNINGS OF CULTURE: A PARADOX

The early human memory appears to have been, like that of a primate, limited to the storage and cued retrieval of specific episodes [9]. Accordingly, Donald [3] uses the term *episodic* to designate a mind such as this that consists only of episodic memories, no abstractions. The episodic mind is dominated by the present moment. Occasionally it encounters a stimulus that is similar enough to some stored episode to evoke a retrieval or reminding event, and sometimes the stimulus evokes a reflexive, or (with much training) learned response. However, it has great difficulty accessing memories independent of environmental cues. It can not manipulate symbols and abstractions, or invent them on its own, and is unable to improve skills through self-cued rehearsal. It seems to encode each episode as a separate, un-modifiable entity.

In contrast, the modern human mind uses abstractions to define *relationships* between episodes, and relates abstractions to one another by way of higher-order abstractions. It can retrieve and recursively operate on memories independent of environmental cues, a process referred to by Karmiloff-Smith [11] as *representational redescription*. By redescribing an episode in terms of what is already known, it gets rooted in the network of understandings that comprise the worldview, and the worldview is perpetually revised as new experiences are assimilated and abstract concepts invented as needed.

The existence of this uniquely human form of cognition leaves us with a nontrivial question of origins. What sort of functional reorganization would turn an episodic mind into that of a modern human? In the absence of representational redescription, how are relationships established such that the memory becomes an interwoven conceptual web? And until a memory incorporates relationships, how can one idea evoke another, which evokes another, in a stream of representational redescription? In other words, if you need a worldview to generate a stream of thought, and streams of thought are necessary to connect memes into a worldview, how could one have come into existence without the other? We have a chicken-and-egg problem.

## 3. AN ANALOGOUS PARADOX: THE ORIGIN OF LIFE

The origin of life presents an analogous paradox: if living things come into existence when other living things give birth to them, how did the first living thing arise? That is, how did something complex enough to reproduce itself come to be? In biology, self-replication is orchestrated through an intricate network of interactions between DNA, RNA, and proteins. DNA contains instructions for how to construct various proteins. Proteins, in turn, both catalyze reactions that orchestrate the decoding of DNA by RNA, and are used to construct a body to house and protect all this self-replication machinery. Once again, we have a chicken-and-egg problem. If proteins are made by decoding DNA, and DNA requires the catalytic action of proteins to be decoded, which came first?

Thus we have two paradoxes—the origin of the psychological mechanisms underlying culture, and the origin of life—which from here on will be referred to as OOC and OOL respectively. The parallels between them are intriguing. In each case we have a system composed of complex, mutually interdependent parts, and it is not obvious how either part could have arisen without the other. In both cases, one part is a storehouse of encoded information about a self in the context of an environment. In the OOL, DNA encodes instructions for the construction of a body that is likely to survive in an environment like that its ancestors survived. In the OOC, an internal model of the world encodes information about the self, the environment, and the relationships between them. In both cases, decoding a segment of this information storehouse generates another class of information unit that coordinates how the storehouse itself gets decoded. Decoding DNA generates proteins that orchestrate the decoding of DNA. Retrieving a memory or concept from the worldview and bringing it into awareness generates an instant of experience, which in turn determines which are the relevant portion(s) of the worldview to be retrieved to generate the *next* instant of experience. (For example, if you had the thought 'my spouse seems sad', you might rack your brain to see what could have caused this.)

## 4. WEAVING CATALYTIC MOLECULES INTO A PRIMITIVE FORM OF LIFE

Kauffman [12] proposed that life may have begun not with a single molecule capable of replicating *itself*, but with a set of *collectively* self-replicating molecules. His proposal combines the concept of organizational closure [13, 14, 17, 18, 23] with insights from random graph theory [4, 5]. When polymers interact, the number of different polymers increases exponentially. However, the number of reactions by which they can interconvert increases faster than their total number. Thus, as their diversity increases, so does the probability that some subset of the total reaches a critical point where there is a catalytic pathway to every member. Such a set is *autocatalytically closed* because each molecule can catalyze the replication of some other molecule in the set, and likewise, its own replication is catalyzed by some other member of the set. (Note that it is *not* necessarily closed in the sense that new molecules cannot be incorporated into the set.) Experimental evidence for this theory using real chemistries [13, 14, 19], and computer simulations [6] have been unequivocally supportive.

## 5. WEAVING MEMORIES INTO A CONCEPTUALLY CLOSED WORLDVIEW

In order for humanity to become capable of evolving culture, the brains of some prehistoric tribe somehow turned into instruments for the variation, selection, and replication of memes. How might Fred, a member of this tribe, have differed from his ancestors such that he was able to initiate this kind of transformation? Donald [3] claims that the transition from episodic to memetic culture "would have required a fundamental change in the way the brain operates." Drawing from the OOL scenario, we will posit that meme evolution begins with the emergence of a collective autocatalytic entity that acts as both code and decoder.

In the OOL case, Kauffman asked: what was lying around on the primitive earth with the potential to act as the 'food set' of a primitive self-replicating system? The most promising candidate is catalytic polymers, the molecular constituents of either protein or RNA. Here we ask: what sort of information unit does the episodic mind have at its disposal? It has memes, specifically memories of episodes. Episodic memes then constitute the food set of our system.

Next, Kauffman asked: what *happens* to the food set to turn it into a self-replicating system? In the OOL case, food set molecules catalyze reactions on each other that increased their joint complexity, eventually transforming some subset of themselves into a collective web for which there existed a catalytic pathway to the formation of each member molecule. An analogous process could transform an episodic mind into a memetic one. Food set memes activate redescriptions of each other that increase their joint complexity, eventually transforming some subset of themselves into a collective web for which there exists a retrieval pathway to the formation of each member meme. Much as polymer A brings polymer B into existence by catalyzing its formation, meme A brings meme B into conscious awareness by retrieving it from memory. Note that a 'retrieval' can be reminding, a redescription of something in light of new contextual information, or a creative blend or reconstruction of many stored memes.

To get more specific about how this might happen, we need to briefly summon what we have learned from neurobiology and cognitive science, and build a best-guess model of human cognition. The first thing to note is that memory is *sparse*. Where $n$ is the number of features the senses can distinguish, N, the number of memes that could potentially be held in attention = $2^n$ for boolean variables (and it is infinitely large for continuous variables). For example, if $n$ =1,000, $N$ = $2^{1,000}$ memes (or even more if we assume that the mind rarely if ever attends all the stimulus dimensions it is capable of detecting). Since assuming $n$ is large, $N$ is enormous, so the number of locations $L$ where memes can be stored is only a small fraction of the $N$ perceivable memes. The number of different memes *actually stored* at a given time, $s$, is constrained by $L$, as well as by the variety of perceptual experience, and the fact that meme retrieval, though distributed at the storage end, is serial at the awareness end. That is, the rate at which streams of thought reorganize the memetic network is limited by the fact that everything is funneled through an awareness/attention mechanism; we can only figure one thing out at a time.

The set of all possible n-dimensional memes a mind is capable of storing can be represented as the set of vertices (if features assume only binary values) or points (if features assume continuous values) in an $n$-dimensional hypercube, where the $s$ stored memes occupy some subset of these points. The distance between two points in this space is a measure of how dissimilar they are, referred to as the Hamming distance. Kanerva [10] makes some astute observations about this space. The number of memes at Hamming distance $d$ away from any given meme is equal to the binomial coefficient of $n$ and $d$, which is well approximated by a Gaussian distribution. Thus if meme X is 111...1 and its antipode is 000...0, and we consider meme X and its antipode to be the 'poles' of the hypersphere, then approximately 68% of the other memes lie within one standard deviation (sqrt[$n$]) of the 'equator' region between these two extremes (Figure 1). As we move through Hamming space away from the equator toward either Meme X or its antipode, the probability of encountering a meme falls off sharply by the proportion sqrt[$n$]/$n$.

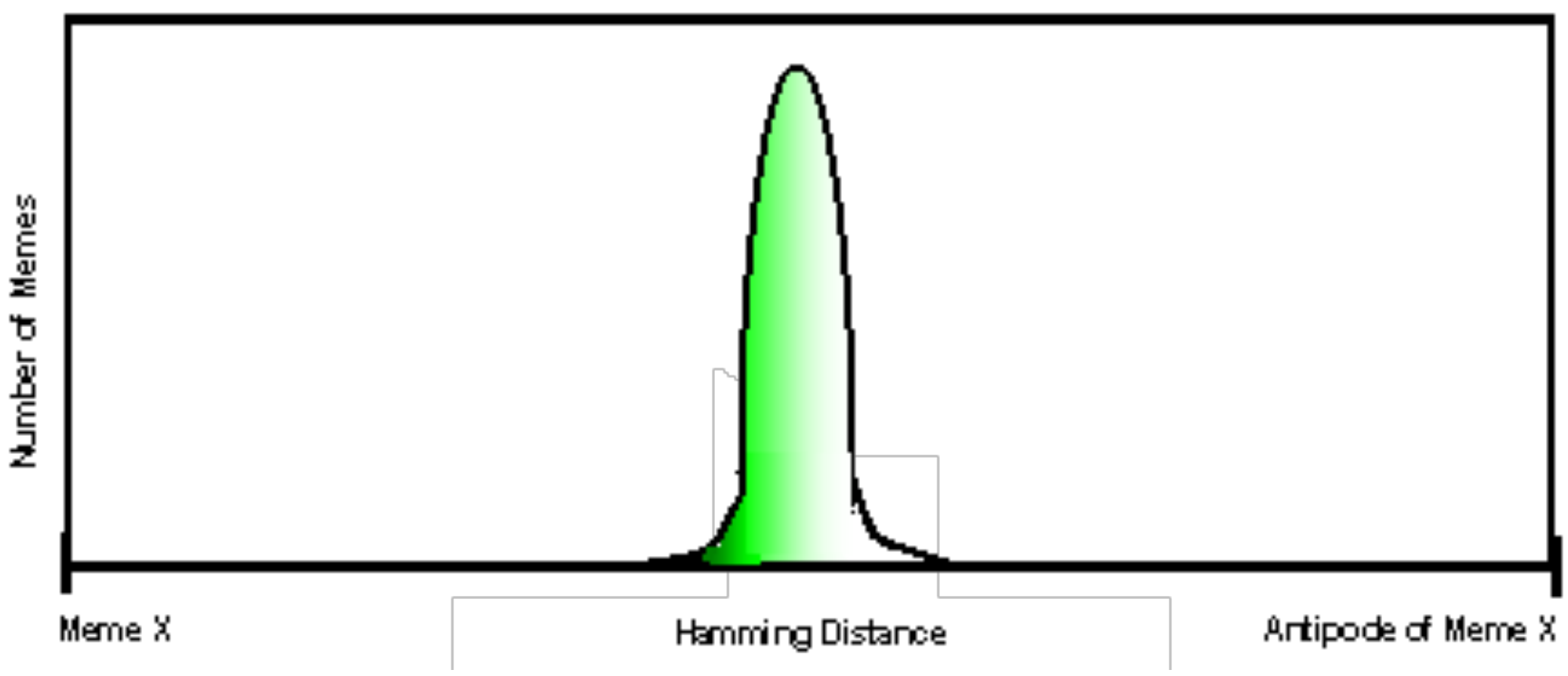

Figure 1. Solid black curve is a schematic distribution of the Hamming distances from address of a given meme to addresses of other memory locations in a sparse memory. The Gaussian distribution arises because there are many more ways of sharing an intermediate number of features than there are of being extremely similar or different. A computer memory stores each item in only the left-most address, whereas a distributed network stores it throughout the network. A restricted activation function, such as the radial basis function, is intermediate between these two extremes. Activation decreases with distance from the ideal address, as indicated by gray shading.

The space of possibilities is so vast that the probability one's current experience is identical to an experience stored in memory is virtually zero. Therefore, retrieval *should* be impossible. In a neural network—a computer architecture inspired by how brains learn and retrieve information—this problem is solved by distributing the storage of a meme across many locations. Likewise, each location participates in the storage of many memes. The stimulus can be represented as input/output nodes, memory locations as hidden nodes, and their pattern of connectivity as weighted links. (It may be that memes are not represented in this sort of granular manner, but the general idea can be adapted to other forms of representation.) An input touches off a pattern of activation which spreads through the network until it relaxes into a stable configuration, or achieves the desired input-output mapping using a learning algorithm. The output vector is determined through linear summation of weighted inputs. Thus, a retrieved meme is not activated from a dormant state, but 'reconstructed', and therefore it is not necessarily identical to the stimulus that evoked it.

How can such a network avoid interference amongst the stored patterns? By *restricting* the distributed activation (as in a radial basis function). In the OOL case, it was crucial that the polymers be catalytic; Kauffman gave each polymer a small, random probability *P* of catalyzing each reaction. Here we do something similar. A hypersphere of locations is activated, such that activation is maximal at the center and tapers off in all directions according to a Gaussian distribution (see Figure 1). The lower the *neuron activation threshold*, the wider this distribution, and therefore the more memes are activated in response to any given meme. Another way the mind prevents interference is by being *modular*; that is, different regions of the brain specialize in the processing of different kinds of information.

The final feature we will note about the brain is that it is *content-addressable*. That is, there is a correspondence between the location in conceptual space where a meme is stored, and its semantic content. Thus each meme can only evoke, or activate, other memes that are similar to it. For example, when considering the problem of having to get out of your car every day to open the garage door, you would not think about doilies or existentialism, but concepts related to the problem—electricity, human laziness, and various openers you have encountered before.

Let us now consider what would happen if, due to a genetic mutation, Fred's neuron activation threshold were significantly lower than average for his tribe. Thus, a greater diversity of memes are activated in response to a given experience, and a larger portion of the contents of memory merge and surface to awareness in the next instant. When meme X goes fishing in memory for meme X', sooner or later this large hypersphere is bound to 'catch' a stored meme that is quite unlike X. For example, since Fred sees the sun every day, there are lots of 'sun-dominated episodes' stored in his brain. Let us say they consist of a sequence of ten 0's followed by a five bit long variable sequence. One night he looks up into the heavens and sees the Evening Star, which gets represented in his focus as 000000011101010. This Evening Star episode will be referred to as meme X. Because the hypersphere is wide, all of the sun memories lie close enough to meme X to get evoked in the construction of X' (as is X itself). Since all the components from which X' is made begin with a string of seven zeros, there is no question that X' also begins with seven zeros. These positions might code for features such as 'appears in sky', 'luminous', *etc*. The following set of three 1s in the 'sun' memes are canceled out by the 0s in the 'Evening Star' meme, so in X' they are represented as *s. These positions might code for features such as 'seen during the day'. The last five bits constituting the variable region are also statistically likely to cancel one another out. These code for other aspects of the experience, such as, say, the smell of food cooking or the sound of wind howling. So X' turns out to be the meme 0000000********, the generic category 'heavenly body', which then gets stored in memory in the next iteration. This evocation of 'heavenly body' by the Evening Star episode isn't much of a stream of thought, and it doesn't bring Fred much closer to an interconnected conceptual web, but it is an important milestone. It is the first time he ever derived a new meme from other memes, his first abstraction, his first creative act.

Once 'heavenly body' has been evoked and stored in memory, the locations involved habituate and become refractory (so, for instance, 'heavenly body' does not recursively evoke 'heavenly body'). However, locations storing memes that have *some* 'heavenly body' features, but that were not involved in the storage of 'heavenly body', are still active. 'Heavenly body' might activate 'moon', and then perhaps 'cloud' *et cetera*, thus strengthening associations between the abstract category and its instances. Other abstractions form in analogous fashion. As Fred accumulates both episodic memes and abstractions, the probability that any given attended meme is similar enough to some previously-stored meme to activate it increases. Therefore reminding acts increase in frequency, and eventually become streams of remindings, which get progressively longer. He is now capable of a train of thought. His memory is no longer just a way-station for coordinating stimuli with action; it is a forum for abstractive operations that emerge through the dynamics of iterative retrieval.

Note that in the OOL case, since short, simple molecules are more abundant and readily-formed than long, complex ones, it made sense to expect that the food set molecules were the shortest and simplest members of the autocatalytic set that eventually formed. Accordingly, in simulations of this process, the 'direction' of novelty generation is outward, joining less complex molecules to form more complex ones through AND operations. In contrast, the memetic food set molecules are complex, consisting of all attended features of an episode. In order for them to form an interconnected web, their interactions tend to move in the opposite direction, starting with relatively complex memes and forming simpler, more abstract ones through OR operations. The net effect of the two is the same: a network emerges, and joint complexity increases. But what this means for the OOC is that there are numerous levels of autocatalytic closure, which convey varying degrees of worldview interconnectedness and consistency on their 'meme hosts'. These levels correspond to increased penetration of the ($n$-1, $n$-2…)-dimensional nested hypercubes implicit in an $n$-dimensional memory space. Since it is difficult to visualize the set of nested, multidimensional hypercubes, we will represent this structure as a set of concentric circles, such the outer skin of this onion-like structure represents the hypercube with all $n$ dimensions, and deeper circles represent lower-dimensional hypercubes (Figure 2). Obviously, not all the nested levels can be shown. The centermost location where a meme is stored is shown as a large, black dot.

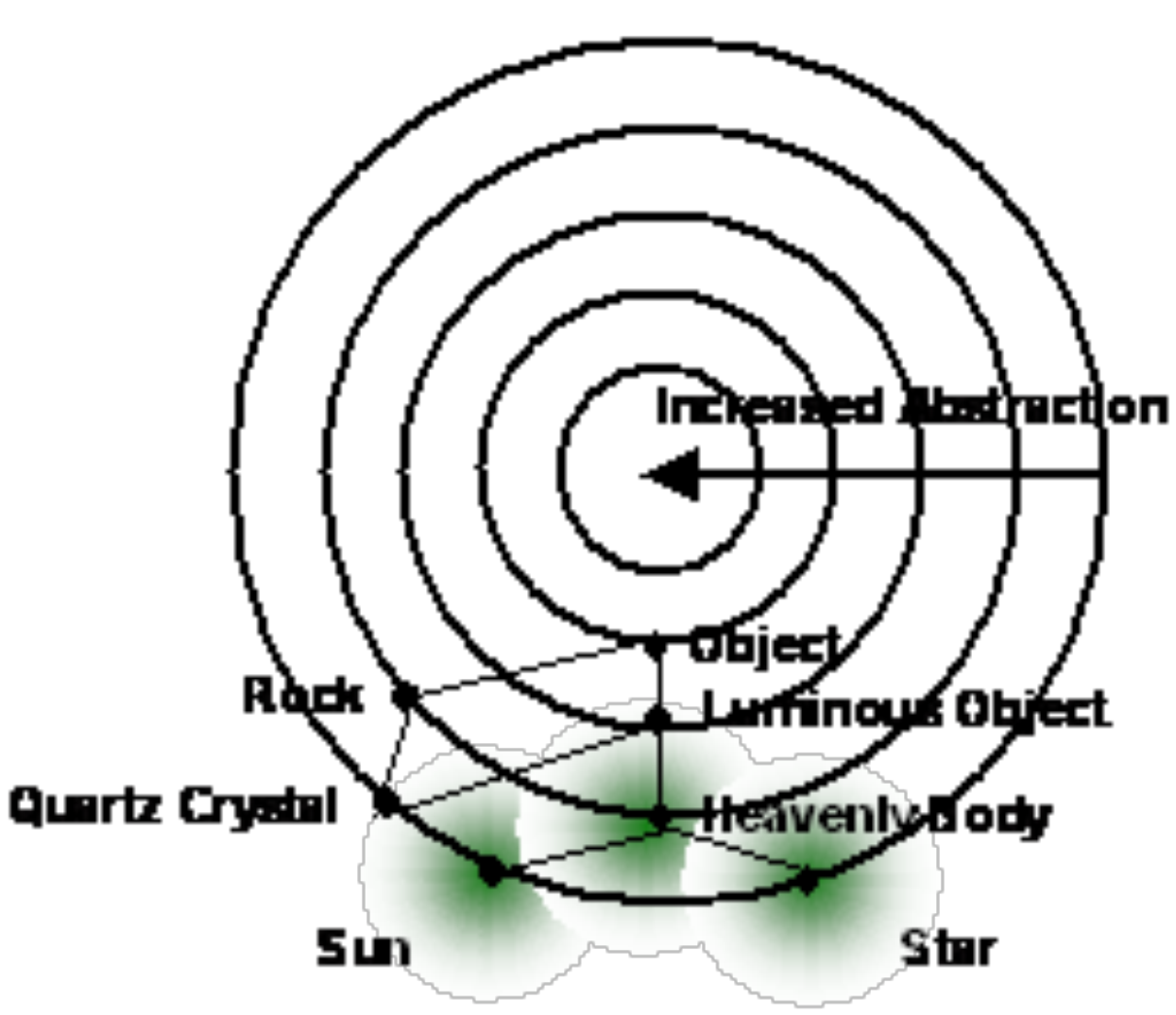

Figure 2. The role of abstractions in creative thought. For ease of visualization, the set of nested hypercubes representing the space of possible memes is shown as a set of concentric circles, where deeper circles store deeper layers of abstraction (lower dimensional hypercubes). A black dot represents the centermost storage location for a specific meme. 'Heavenly body' is a more general concept than 'sun' or 'star', and is therefore stored at a deeper layer of abstraction. Grey circle around each stored meme represents hypersphere where the meme gets stored and from which the next meme is retrieved.

The outermost shell encodes memes in whatever form they are in the first time they are consciously encountered. This is all the episodic mind has to work with. In order for one meme in this shell to evoke another, they have to be extremely similar at a superficial level. In a memetic mind, however, related concepts are within reach of one another because they are stored in overlapping hyperspheres. 'Sun' and 'star' might be too far apart in Hamming distance for one to evoke the other directly. However, by attending the abstraction 'Heavenly Body', which ignores the 'seen at night versus seen during the day' distinction, the memetic mind decreases the apparent Hamming distance between them.

Under what conditions will the transformation from discrete memories to interconnected conceptual web actually occur? In the OOL case, Kauffman had to show that R, the number of reactions, increased faster than N, the number of polymers; thus for a large range of values of R and N, the system inevitably reaches a phase transition to a critical state wherein for some subset of memes there exists a retrieval pathway to each meme in the subset. How do we know that streams of thought will do the same thing? We need to show that some subset of the memes stored in an individual's mind inevitably reach a critical point where there exists a retrieval pathway by which every meme in that subset can get evoked. But here, it is *not* reasonable to assume that all $N$ perceivable memes actually exist (and can therefore partake in retrieval operations). The awareness/attention filter presents a bottleneck that has no analog in the OOL scenario. As a result, whereas OOL polymers underwent a *sharp* transition to a state of autocatalytic closure, the transition in inter-meme relatedness is expected to take place *gradually*. So we need to show that $R$, the diversity of ways one meme can evoke another, increases faster than not $N$ but $s$, the number of stored memes (i.e., memes that have made it through this bottleneck). That is, as the memory assimilates memes, it comes to have more ways of generating memes than the number of memes that have explicitly been stored in it.

Under what conditions does $R$ increases faster than $s$? The reader is referred to [8] for the mathematical details, but the key idea is that abstraction increases $s$ by creating a new meme, but it increases $R$ more, because the more abstract the concept, the greater the number of memes a short Hamming distance away (since irrelevant dimensions make no contribution to Hamming distance). Moreover, as $n$ starts to decrease the number of possible abstractions for each value of $n$ increases (up to $M/2$, after which it starts to decrease). Whereas $R$ increases as abstraction makes relationships amongst memes increasingly explicit, $s$ levels off as new experiences have to be increasingly unusual in order to count as new and get stored in a new constellation of locations. Furthermore, when the carrying capacity of the memory is reached, $s$ plateaus, but $R$ does not. Thus, as long as the neuron activation threshold is large enough to permit abstraction and small enough to permit temporal continuity, the average value of $n$ decreases, and sooner or later, the system is expected to reach a critical percolation threshold such that $R$ increases exponentially faster than $s$. The memory becomes so densely packed that any meme that comes to occupy the focus is bound to be close enough in Hamming distance to some previously-stored meme(s) to evoke it. The memory (or some portion of it) is holographic, in the sense that there is a pathway of associations from any one meme to any other; together they form an autocatalytic set. What was once just a collection of isolated memories is now a structured network of concepts, instances, and relationships—a worldview.

Now that we have an autocatalytic network of memes, how does it self-replicate? In the OOL scenario, polymer molecules accumulate one by one until there are at least two copies of each, and their shell divides through budding to create a second replicant. In the OOC scenario, Fred shares concepts, ideas, stories, and experiences with his children and tribe members, spreading his worldview meme by meme. In Farmer et al.'s OOL simulation [6], mentioned earlier, the probability of autocatalysis could be increased by raising either the probability of catalysis or the number of polymers. Something similar happens here. Even if Fred's daughter Pebbles has a higher neuron activation threshold than Fred, once she has assimilated enough of Fred's abstractions, her memes become so densely packed that a version of Fred's worldview snaps into place in her mind. Pebbles shares *her* worldview with her friend Bambam, who in turn shares it with the rest of the tribe. These different hosts expose their 'copy' of Fred's original worldview to different experiences, different bodily constraints, sculpting them into unique internal models of the world. Small differences are amplified through positive feedback, transforming the space of viable worldview niches. Individuals whose activation threshold is too small to achieve worldview closure are at a reproductive disadvantage, and, over time, eliminated from the population. Eventually the proclivity for an ongoing stream of thought becomes so firmly entrenched that it takes devoted yogis years of meditation to even briefly arrest it.

## 6. WORLDVIEW EXPANSION AND VIABILITY

The most primitive level of autocatalytic closure is achieved when stored episodes are interconnected by way of abstractions just a few 'onionskin layers' deep, and streams of thought zigzag between these superficial layers. A second level occurs when relationships amongst *these* abstractions are identified by higher-order abstractions at deeper onionskin layers. *Et cetera*. Once an individual has defined an abstraction, identified its instances, and chunked them together in memory, she can manipulate the abstraction much as she would a concrete episode.

Categorization creates new lower-dimension memes, which makes the space denser, and increases susceptibility to the autocatalytic state. On the other hand, creating new memes by combining stored memes could interfere with the establishment of a sustained stream of thought by decreasing the modularity of the space, and thereby decreasing density. If cross-category blending indeed disrupts conceptual closure, one might expect it to be less evident in younger children than in older ones, and this expectation is born out experimentally [11]. There is evidence of a similar shift in human history from an emphasis on ritual and memorization toward an emphasis on innovation [3]. As world-views become more complex, the artifacts we put into the world become more complex, which necessitates even more complex world-views, et cetera, thus a positive feedback cycle sets in.

Van de Vijver [22] suggests that the cognitive system grows through a process of identification, a view that is highly compatible with the ideas proposed here. It is attended stimuli that get assimilated and integrated into the worldview, and attention involves an identity relationship between the stimulus, and the mental representation of it. Not necessarily all encountered stimuli would be expected to undergo this identification process. Much as biological closure shields off toxic substances yet promotes the assimilation of food necessary for maintenance and growth, conceptual closure involves censorship of potentially harmful memes, and assimilation of ones that could generate thought trajectories that enhance individual wellbeing (in other words, the fruits of their travels in conceptual space manifest as enhanced contextuality in physical space). The more stimuli the conceptual system has assimilated, the more of external reality it can capture, thus the greater the potential of the individual to adapt itself to its environment. However, accuracy is not the only determinant of what makes a successful worldview. A viable worldview is one that reinforces thought trajectories that lead to behaviors that enhance individual wellbeing. As Trivers [20, 21] suggests, this may involve a certain amount of bias or self-deception.

In fact, though it would seem that in the transition from the episodic mind to the memetic mind, we have made enormous progress, this isn't necessarily the case. The episodic mind, in fact, has *no* inconsistencies. Since it doesn't represent relationships, it doesn't get any relationships wrong. It never encoded the sun and the stars as different kinds of entities in the first place, so if it were to go further and further away from the sun until it realized that the sun is just like any other star, no conceptual adjustment would have to be made. Thus there is a tradeoff between abstraction and accuracy.

Once a new stimulus or episode has been attended, it still has to get incorporated into the conceptual network; its relationships to previously identified stimuli need to be worked out. Kauffman suggests that the autocatalytic molecular set oscillates between a *supracritical phase*—wherein the system is robust enough to accept new molecules—and a *subcritical phase*—wherein the integration of recently-acquired molecules temporarily challenges the system's robustness, such that new ones are not accepted. In other words, the system grows by cycling back and forth between an open state wherein new inputs are acquired, and a closed state wherein these inputs are integrated into the system at large. It may be that the sleep-wake cycle is an analogous oscillation of the conceptual system between perceptual openness (awake), and integration of perceived stimuli (sleep). During the day, the parent may help keep the child's mind perpetually poised at a supracritical state by interacting with the child in ways that promote the formation of novel abstractions. This kind of parental guidance is analogous to handcrafting new polymers to be readily-integrated into a particular autocatalytic set.

We began by noting that conceptual closure increases the potential contextuality of a system's behavior. We generally use contextuality as a rough indicator of degree of awareness; the more contextual the system's behavior, the more likely we are to say that it is conscious, that there is something it is like to *be* that system. This suggests that each successive level of closure—first biological, then conceptual—locks the system into a more concentrated state of awareness, like light trapped in mirrors. This would require that there exist some kind of nascent, proto-awareness prior to closure, which seems unlikely. However the idea is not unpopular, even in academia. According to Chalmers' [1] double aspect theory, for at least some information spaces, whenever an information state in that space is realized physically, it is also realized phenomenally. (An extreme version is panpsychism—the idea that all information has a conscious aspect.) The theory seems very counterintuitive. However, if it *were* true, it is unlikely we would be aware of it. Why? Because we are the products of millions of years of selective forces perfecting our ability to accentuate the subjectivity we ourselves experience, and shield off the subjectivity other entities are experiencing. This is necessary in order to convince a living system to consume other plants or animals in order to survive. (If an entity valued all conscious experience equally, it would not act to preserve its own experience at the expense of the experience of other entities.) Thus, the double aspect theory of information is consistent with the possibility that closure is accompanied by a phase shift in the potential for conscious experience.